\documentclass[
superscriptaddress,
preprintnumbers,
amsmath,amssymb,
aps,
prb,
twocolumn,
floatfix,
]{revtex4-2}
\usepackage{natbib}
\usepackage{graphicx}
\usepackage{dcolumn}
\usepackage{bm}
\usepackage{subcaption}
\usepackage{xcolor}

\begin{document}
\author{Gen Li}
\affiliation{School of Physics, Institute for Quantum Science and Engineering, and Wuhan National High Magnetic Field Center, Huazhong University of Science and Technology, Wuhan 430074, China}
\author{Bing-Zhong Hu}
\affiliation{School of Physics, Institute for Quantum Science and Engineering, and Wuhan National High Magnetic Field Center, Huazhong University of Science and Technology, Wuhan 430074, China}
\author{Wen-Hao Mao}
\affiliation{School of Physics, Institute for Quantum Science and Engineering, and Wuhan National High Magnetic Field Center, Huazhong University of Science and Technology, Wuhan 430074, China}
\author{Nuo Yang}
\affiliation{State Key Laboratory of Cool Combustion, and School of Energy and Power Engineering, Huazhong University of Science and Technology, Wuhan 430074, China}
\author{Jing-Tao L\"u}
\email{jtlu@hust.edu.cn}
\affiliation{School of Physics, Institute for Quantum Science and Engineering, and Wuhan National High Magnetic Field Center, Huazhong University of Science and Technology, Wuhan 430074, China}

\title{Boosting current-induced molecular dynamics with machine-learning potential}

\date{\today}
\begin{abstract}
In a current-carrying single-molecular junction (SMJ), a hierarchy of hybrid energy transport processes takes place under a highly nonequilibrium situation, including energy transfer from electrons to molecular vibrations via electron-vibration interaction, energy redistribution within different vibrational modes via anharmonic coupling, and eventual energy transport to surrounding electrodes. A comprehensive understanding of such processes is a prerequisite for their potential applications as single-molecular devices. $Ab$ $initio$ current-induced molecular dynamics (MD) is an ideal approach to address this complicated problem. But the computational cost hinders its usage in systematic study of realistic SMJs. Here, we achieve orders of magnitude improvement in the speed of MD simulation by employing machine-learning potential with accuracy comparable to density functional theory. Using this approach, we show that SMJs with graphene electrodes generate order of magnitude less heating than those with gold electrodes. Our work illustrates the superior heat transport property of graphene as electrodes for SMJs, thanks to its better phonon spectral overlap with molecular vibrations.
\end{abstract}
\maketitle

\section{INTRODUCTION}
Recent development of scanning thermal probe technique has enabled experimental measurement of thermal conductance across single-molecular junctions (SMJs) and metallic atomic chain junctions\cite{Cui2017,Mosso2017,cui_thermal_2019,Mosso2019,Mosso2019a}, one and two decades later than the first thermoelectric\cite{Reed1997} and electrical\cite{Reddy2007} measurement, respectively. This opens exciting opportunities for systematic investigation of thermal transport in atomic-scale junctions, i.e., through chemical engineering\cite{xiang_molecular-scale_2016,Cui2017a,xin_concepts_2019,Evers2019,Segal2016,Gotsmann2022,Segal2003,MoghaddasiFereidani2019,Klockner2017,Klockner2016,Li2015,Li2017}. 

A more challenging problem is nonequilibrium thermal transport between electronic and vibrational/phononic degrees of freedom in the presence of flowing electrical current\cite{Galperin2007}. 
Wherein, two important bottleneck processes controlled by electron-vibration/phonon and anharmonic phonon interaction are simultaneously at play. 
First, electrons transfer energy to the molecular vibrations via electron-vibration interaction. Second, the excess energy is redistributed within molecular vibrational modes and further transported to the surrounding electrodes, in which anharmonic vibrational coupling may play a key role. 
For a comprehensive understanding, all these processes need to be addressed on an equal footing. Theoretical studies of the former have revealed at least two energy transfer mechanisms\cite{Dundas2009,Lu2010}: stochastic Joule heating and work performed by deterministic non-conservative current-induced force. While Joule heating is always present, the non-conservative current-induced force becomes important in high conductance SMJs. Experimentally, the signature of Joule heating has been probed by electrical and optical methods\cite{Huang2006,Ward2011,Ioffe2008,Tsutsui2008}, while only preliminary evidence of current-induced non-conservative force has been reported\cite{Sabater2015}.  
On the other hand, studies on vibrational energy redistribution within the molecule and subsequent energy transport to the electrodes are scarce. Theory has been developed to account for electron-vibration and anharmonic vibrational coupling simultaneously. However, combining with first-principles calculations to study realistic SMJs is numerically challenging.  
In a previous report, $ab$-$initio$ current-induced molecular dynamics  (MD) simulation has been performed for prototypical molecular junctions to study the atomic scale hot-spot and energy redistribution\cite{Lu2020}. However, the computational cost prevents systematic study of different types of junctions.

In this work, we overcome this difficulty by employing machine-learning potential derived from $ab$-$initio$ MD data. We perform a comparative study on heat transport property of alkane and carbon chain junctions with gold and graphene electrodes. We show that good spectral overlap between molecular vibrations and graphene phonons ensures effective harmonic energy transport between them and results in much less heating of the molecule as compared to gold electrodes. Our work illustrates the superior advantage of graphene as electrodes for constructing stable SMJs\cite{jia_covalently_2016,Gutierrez2019,LiGen2021}.

\begin{figure}
	\centering
    \includegraphics[width=.75\columnwidth]{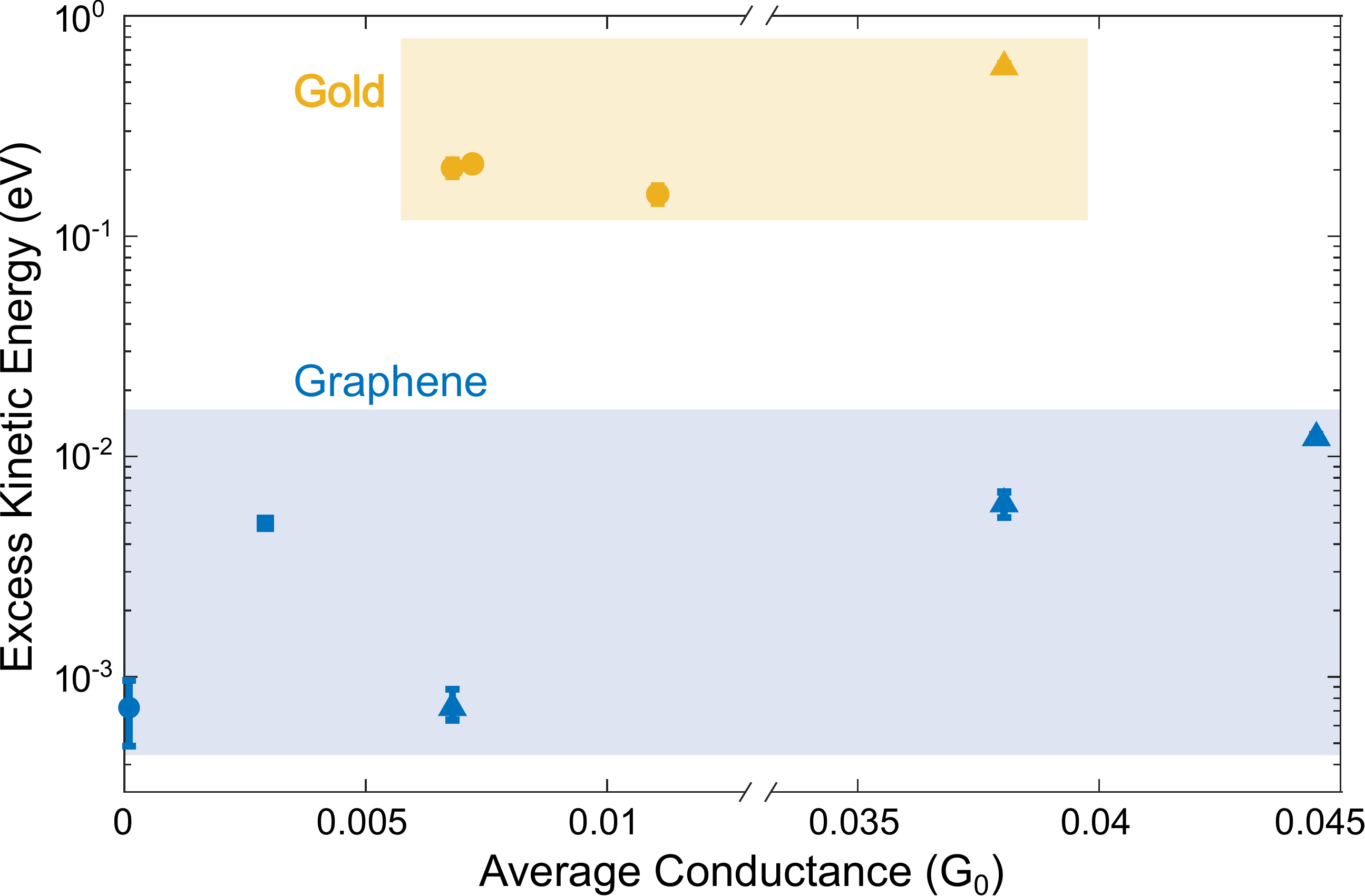}
    \caption{Excess kinetic energy of central molecule for junctions with different conductance between gold (yellow) and graphene (blue) electrodes at a voltage bias  $V= 1$ V. Heating of gold junctions is $\sim 1 - 2$ orders of magnitude larger than that of graphene junction. The circle points are results from alkane (CH$_2$) structures, triangle and square points are from carbon chain (CH) structures (Fig. S2).}
    \label{fig:comp}
\end{figure}

\section{THEORY} 
Our theory is based on a semiclassical Langevin equation\cite{Lu2019,kantorovich_generalized_2008,kantorovich_generalized_2008-1,wang_quantum_2007,Dou2017}
\begin{equation}
\ddot{\mathbf{Q}} = -\nabla_{\textbf{Q}}V - {\bf \Gamma} \dot{\textbf{Q}} - {\bf A} {\bf Q} + \textbf{f}.
\end{equation}
The vector ${\bf Q}$ is mass-normalized displacement vector containing all degrees of freedom of the central extended molecule, which can also include some electrode atoms. The first term at the right hand side accounts for the total potential force exerted on the extended molecule, the second and third terms are the friction and renormalization terms, and the final term is the fluctuating force. The last three terms include contributions from three baths:  two phonon baths representing left and right electrodes, and one electron bath in nonequilibrium steady state with flowing electrical current (Fig.~\ref{fig:excitation} insets), i.e., for the friction matrix ${\bf \Gamma}={\bf \Gamma}_{\rm L}+{\bf \Gamma}_{\rm R}+{\bf \Gamma}_{\rm e}$.

In the numerical calculation, we make use of the relation between the friction matrix and the self-energies in the nonequilibrium Green's function (NEGF) theory\cite{Lu2012}, i.e., in frequency domain, 
\begin{equation}
   {\bf \Gamma}(\omega) = -{\rm Im}{\bf \Pi}^r(\omega)/\omega, \quad  {\bf A}(\omega) = {\rm Re}{\bf \Pi}^r(\omega). 
\end{equation}
The fluctuating force ${\mathbf f}$ is characterized by the time correlation function\cite{Lu2012,Lu2019}
\begin{equation}
    \langle {\bf f}_{\mu}(t) {\bf f}^T_{\nu}(t') \rangle =\frac{i}{2} \hbar \delta_{\mu,\nu} [{\bf \Pi}_\mu^>(t-t')+{\bf \Pi}_\mu^<(t-t')], \quad  \mu, \nu = {\rm L}, {\rm R}, {\rm e}.
\end{equation}
Here, ${\bf \Pi}^r$, ${\bf \Pi}^{>}$, ${\bf \Pi}^<$ are the retarded, greater, and less self-energies in the NEGF theory\cite{Lu2012,Lu2019}, with ${\bf \Pi}^r={\bf \Pi}^r_{\rm e}+{\bf \Pi}^r_{\rm R}+{\bf \Pi}^r_{\rm L}$. Notably, since the electron bath is in nonequilibrium steady state, the fluctuating force includes standard thermal, quantum zero-point fluctuations, and extra nonequilibrium contribution due to flowing electrical current. The nonequilibrium fluctuating force breaks the fluctuation-dissipation relation, and is responsible for Joule heating of the junction.  Moreover, under nonequilibrium situation, ${\bf A}$ may become asymmetric. Its anti-symmetric part represents the nonconservative current-induced force. 

A schematic diagram of our numerical procedure is depicted in Fig.~S1. First-principles calculation at the density functional theory (DFT) level is used to calculate the electronic structure, molecular vibrational modes, electron-vibration interaction within the molecule, based on which the NEGF self-energies are evaluated. Here the SIESTA-TranSIESTA-Inelastica toolkit is used for this purpose\cite{Soler2002,Brandbyge2002,Frederiksen07}. In addition, the wideband approximation is used when calculating the friction matrix\cite{Lu2012}. In the MD simulation, the atoms of outermost layers are fixed, and the rest are dynamical. The coupling of the three baths to the molecular system is shown in the insets of  Fig.~\ref{fig:excitation}. The electron bath couples only to the molecule (under red lines). The left and right phonon baths couple instead to the system atoms under the two blue lines. Periodic boundary condition is used in the direction perpendicular to the transport direction. The coupling to each phonon bath is approximated by one diagonal friction matrix with the same diagonal element $\gamma_{\rm L/R} = 10$ ps$^{-1}$. The accuracy of this approximation can be improved by including more electrode atoms in the extended molecule to account for non-Markovian memory effect\cite{Stella2014,ness_applications_2015}.

MD simulation is performed using machine-learning potentials trained by $ab$-$initio$ MD data of the same structure using DeePMD-kit\cite{wang_deepmd-kit_2018}.
The obtained potential can accurately reproduce the vibrational modes of each structure from DFT (Fig.~S4). 
Without losing numerical accuracy, we can perform MD simulation with a speed 1-2 orders of magnitude faster than standard $ab$-$initio$ MD. This is the most important step forward compared to our previous work\cite{Lu2020}. 
Results shown below are obtained by averaging three independent MD runs, where each run lasts for 4 ns, with a MD time step $dt=0.5$ fs. Such calculations are not feasible for $ab$ $initio$ MD due to the expensive computational cost.

\section{RESULTS AND DISCUSSIONS}
\begin{figure*}
	\centering
    \includegraphics[width=1.8\columnwidth]{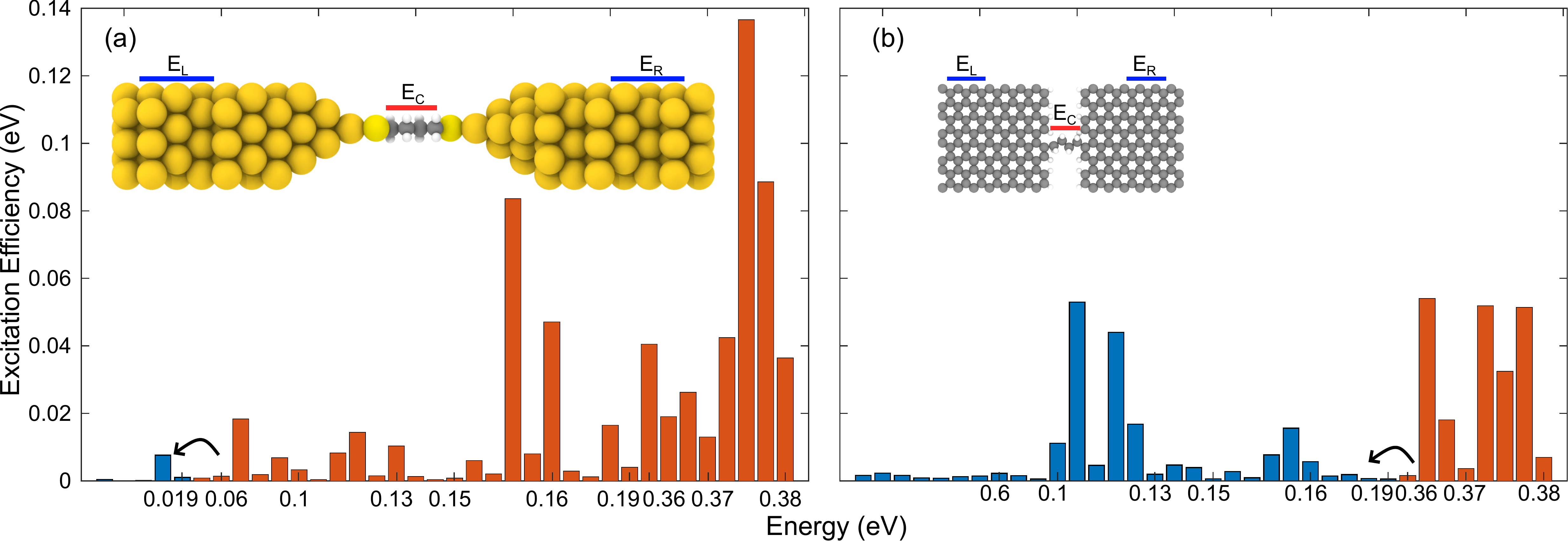}
    \caption{Electronic excitation efficiency of molecular vibrational modes [Eq.~(\ref{eq:excitation})] for gold (a) and graphene (b) junctions. Blue modes have direct frequency overlap with electrode phonons. Energy transport through these modes to surrounding electrodes is mainly harmonic. Meanwhile, red modes have frequencies higher than the phonon band width. Thus, anharmonic frequency down-conversion is necessary before their excess energy is transported to surrounding electrodes  (arrows). Insets: Alkane chain consisting of 4 CH$_2$ units placed between gold and graphene electrodes. Alkane connects to gold through thiol bonds Au-S-C and to graphene through direct C-C bonds. 
    OVITO is used for visualization.\cite{ovito}}
    \label{fig:excitation}
\end{figure*}

Our main results are summarized in Fig.~\ref{fig:comp}, where alkane (CH$_2$ unit) and hydrogenated carbon-chain (CH unit) junctions with gold and graphene electrodes are considered at a voltage bias of $1$ V. Their structures are shown in the insets of Fig.~\ref{fig:excitation} and in Fig.~S{2}, with the corresponding electrical transmission in Fig.~S{3}. We have considered junctions with a wide range of electrical conductance by manually shifting the average chemical potential or by changing the molecular unit from $sp^3$ CH$_2$ in alkane to $sp^2$ CH in carbon chain. 
We characterize heating of the molecule by its excess kinetic energy $\Delta E_K$, which is the difference of  molecular kinetic energy at $1$ V and $0$ V. We observe order of magnitude more heating in Au junction compared to graphene junction, independent of the type of molecule or electrical conductance. 
In the following, we show that less heating of the graphene junction is rooted in the better spectral overlap between molecular vibrations and graphene phonons. This results in efficient harmonic energy transport to surrounding electrodes and less heating of the molecule.

To quantify the electronic excitation of each vibrational mode $n$, we define an excitation efficiency (Supporting Information (SI), Sec. I-1)
\begin{equation}
\gamma_n = \hbar\omega_n \frac{\chi^+_{nn}}{\eta_{nn}}.
\label{eq:excitation}
\end{equation}
Here, $\hbar\omega_n$ is the energy of mode $n$, $\chi^+_{nn}$ and $\eta_{nn}$ characterize the nonequilibrium excitation and damping of the vibrational mode due to coupling to nonequilibrium electrons.

Figure~\ref{fig:excitation} shows excitation efficiency of alkane junctions with gold (a) and graphene (b) electrodes. Modes with blue color lie within the phonon band width of the electrodes, while those with red color are out of the band width.  
Harmonic phonon energy transport from the blue modes to the two electrodes is the most effective channel to conduct excess energy out of the molecule (harmonic channel). Excess energy stored in red modes has to be down-converted into lower frequency modes via anharmonic interaction before being transferred to electrodes (anharmonic channel). For gold junction, the harmonic channel is much less efficient, since vibrational modes that interact strongly with electrons are all out of the energy range of harmonic channel. Meanwhile, for graphene junction, only C-H stretching modes ($> 0.3$ eV) are out of the graphene phonon band. Thus, effective harmonic energy transport is expected in graphene junction.

\begin{figure}
	\centering
    \includegraphics[width=.92\columnwidth]{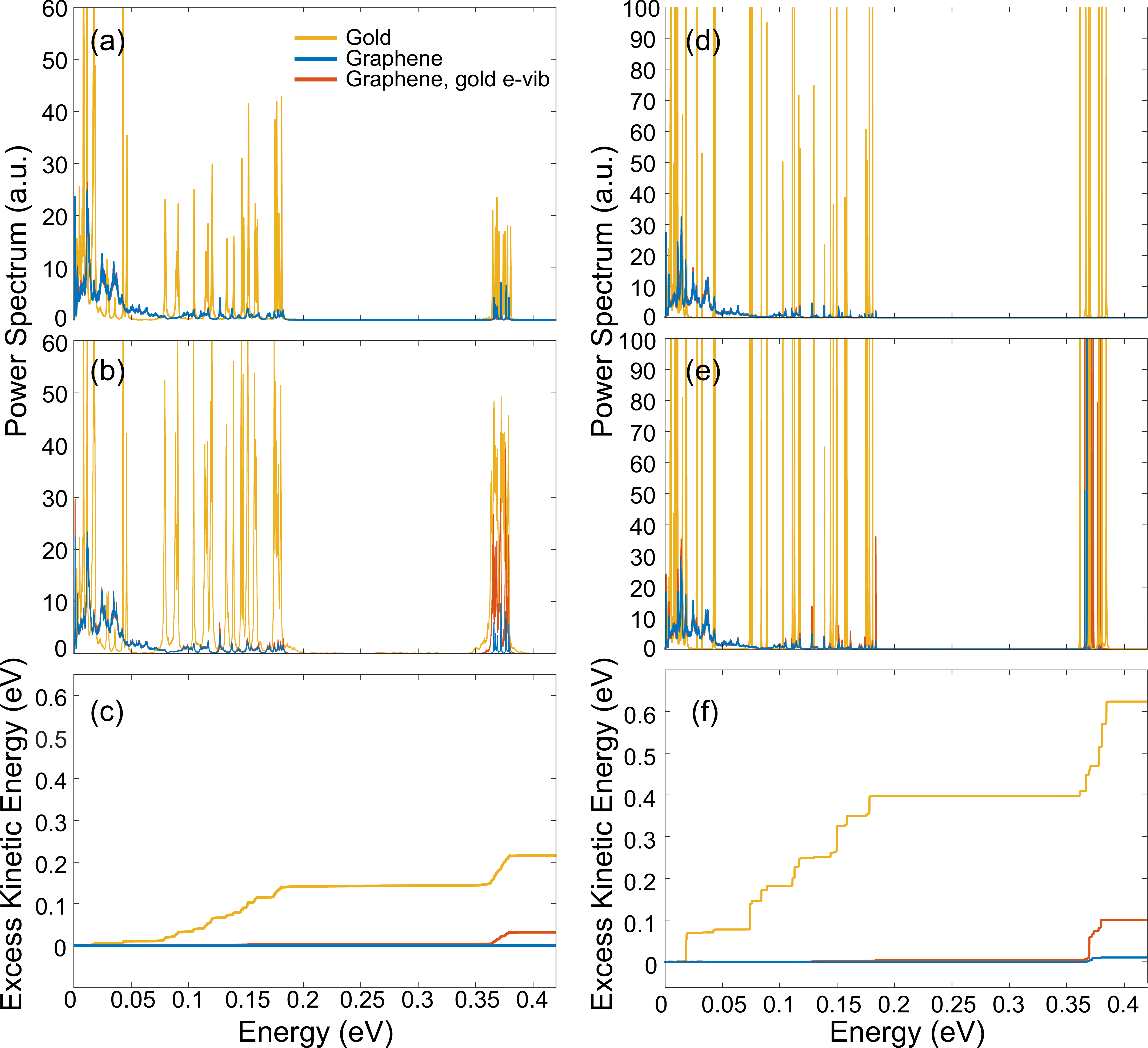}
    \caption{Kinetic energy power spectra $C_{vv}(\omega)$ [Eqs.~(S5-S7)] of the molecule at V= 0 V (a, d), V= 1 V (b, e), and  the corresponding cumulative integration of the excess kinetic energy $\Delta E_K(\omega) = \int_0^\omega d\omega' [C_{vv}(\omega', 1 V)-C_{vv}(\omega', 0 V)]$ (c, f) of the central molecule between gold and graphene electrodes. (a-c) and (d-f) are results obtained from the machine-learning potential and the harmonic potential, respectively. Yellow and blue lines are results obtained from gold and graphene electrodes, respectively. The red lines are results of graphene electrodes but with electron-vibration interaction from gold junctions. To get larger current, we have shifted the Fermi level of graphene electrodes to $0.3$ eV as shown in Fig.~S3. In experiments, this can be achieved by electrostatic gating.} 
    \label{fig:power}
\end{figure}

The above analysis is confirmed by analyzing the kinetic energy power spectra (SI, Sec. I-2) of the molecule obtained from MD simulation in Fig.~\ref{fig:power}. 

\begin{figure*}
	\centering
	\includegraphics[width=1.8\columnwidth]{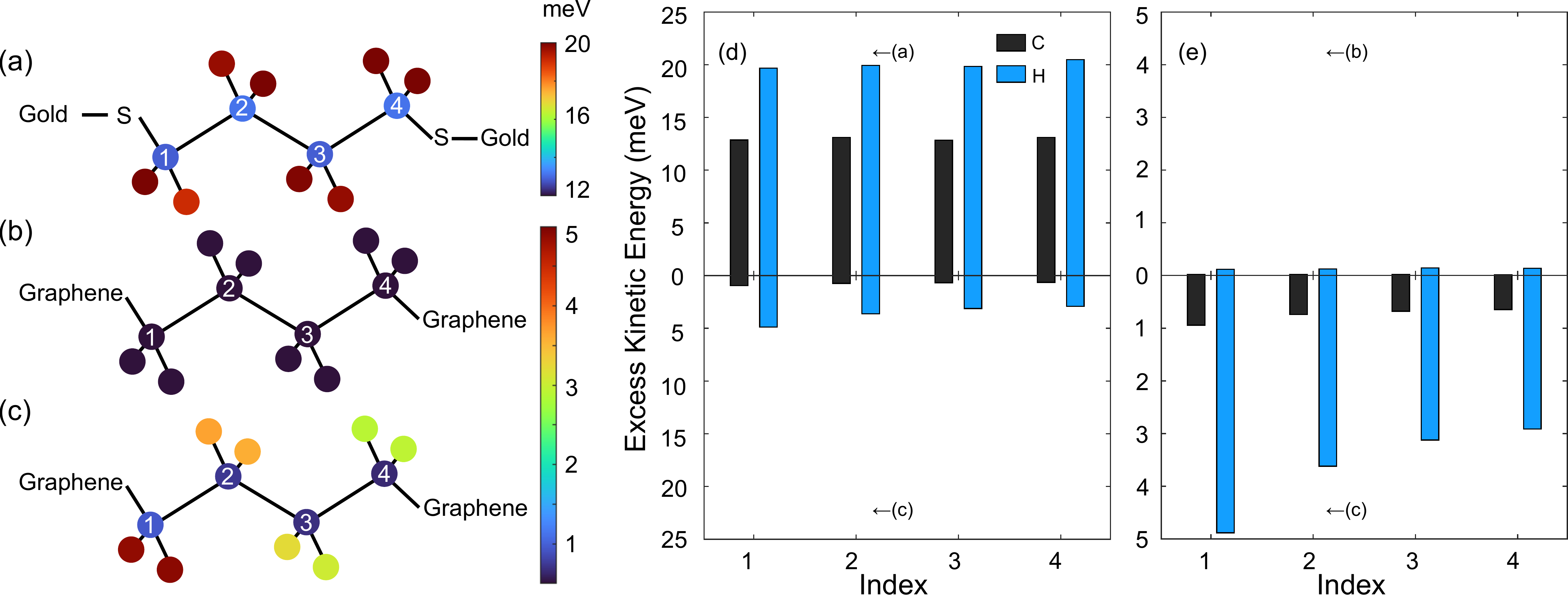}
    \caption{Real space excess kinetic energy distribution within the molecule for different electrodes and electron-vibration coupling. (a) and (b) show results for gold and graphene electrodes, respectively. (c) The phonon properties are taken to be the same as (b), but the electron-vibration coupling is taken the same as (a). (d) Comparison between (a) and (c). (e) Comparison between (b) and (c). Note the different scales of $y$ axis.}
    \label{fig:ker}
\end{figure*}

The power spectra within the electrode phonon bandwidth show obvious broadening due to coupling to electrode phonons, which is within [0, 0.2] eV for graphene electrodes, while only within [0,0.02] eV for gold electrodes. 
Better spectral overlap between molecular vibration and electrode phonons results in only observable heating of the C-H bond ($> 0.3$ eV) for graphene junction. The situation is drastically different for gold junction, where heating in a much wider frequency range is observed. This is consistent with Fig.~\ref{fig:excitation}. The harmonic channel is much less effective in gold junction, since the vibrational modes that overlap with gold phonon band is much less excited than those out of the phonon band. 

\begin{figure}
	\centering
    \includegraphics[width=0.99\columnwidth]{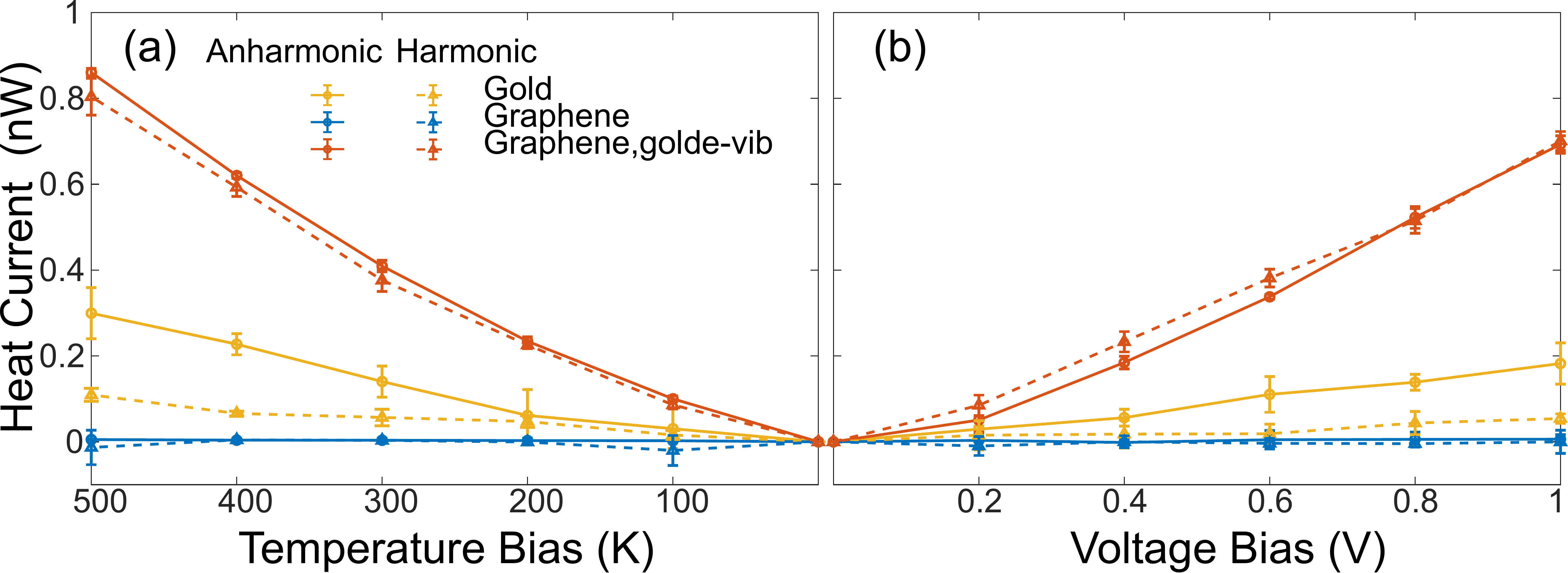}
    \caption{Heat current as a function of temperature bias applied between electron and phonon systems (a) and voltage bias applied between two electrodes (b). Circles are obtained from anharmonic SCLMD using machine-learning potentials, while triangles are obtained from harmonic calculation. The color code is the same as Fig.~\ref{fig:power}.}
    \label{fig:hf}
\end{figure}

The steady state spectral distribution is a result of balance between energy input from the nonequilibrium electron bath and output to two surrounding phonon baths. 
To further dis-entangle these two effects, we have performed extra simulation of the graphene junction using electron coupling parameters from gold junction. The results are shown as red curves in Fig.~\ref{fig:power} (see also Fig.~\ref{fig:ker} (c-e) and Fig.~\ref{fig:hf}). More heating and larger heat current (Fig.~\ref{fig:hf}) can be observed in this case. But, the excess energy is still much smaller than the gold junction. Thus, we conclude that smaller heating in graphene junction is mainly due to better frequency overlap of the molecular vibrations with two phonon baths. 
The effect of anharmonic channel can be deduced by comparing results using the anharmonic machine-learning potential [Fig.~\ref{fig:power}(a-c)] and the harmonic approximation [Fig.~\ref{fig:power}(d-f)]. The excess kinetic energy drops more than 50\% for both junctions when including anharmonic vibrational coupling. 

Excess heating of the molecule can also be studied within real space (Fig.~\ref{fig:ker}). Previous studies have reported asymmetric energy distribution in metal atomic chain (atomic hot-spot) originated from deterministic energy transfer of non-conservative current-induced force\cite{lu2015}. Normally, this requires that the junction has a high conductance ($\sim 1$ G$_0$). But the conductance of the molecular junctions we consider here is much lower. The effect of the non-conservative force is negligible and we obtain quite symmetric energy distribution. The asymmetry distribution of H atoms in Fig.~\ref{fig:ker} (c)  is possibly due to the structure asymmetry, instead of the effect of current-induced forces. 

We now turn to the heat current
from electron to molecular vibrations (Fig.~\ref{fig:hf}). In Fig.~\ref{fig:hf}(a)  temperature of electron bath increases from 300 K up to 800 K at zero voltage bias while keeping the phonon bath temperature fixed at 300 K. In Fig.~\ref{fig:hf}(b), voltage bias is applied  to electrons between left and right electrodes while keeping the temperature at $300$ K. A temperature bias of $\sim 400$ K is needed to generate similar heat current at $V = 1$ V. Comparing the two types of junctions, anharmonic coupling is more important for gold electrodes. Heat current increases by $> 100\%$ when including the anharmonic coupling. While for graphene electrodes, the harmonic and anharmonic results differ only slightly, indicating negligible role of anharmonic effect. The difference lies again in the spectral overlap between molecule and electrodes. 

\section{CONCLUSION}
In conclusion, with the help of machine learning potential with accuracy comparable to density functional theory, we are able to perform a systematic study on current-induced energy transport between electrons and phonons in SMJs. We show that better spectral overlap between electrode phonons and molecular vibrations makes graphene electrodes promising candidate for developing stable single-molecular devices. Our work paves the way for studying current-induced molecular dynamics\cite{Schirm2013,Preis2021} and nonequilibrium heat transport between electrons and phonons in realistic SMJs with reasonable computational cost.

\begin{acknowledgements}
This work was supported by the National Natural Science Foundation of China (No. 21873033).
We thank the National Supercomputing Center in Shanghai for providing computational resources.
\end{acknowledgements}


\appendix
\setcounter{figure}{0}
\setcounter{equation}{0}
\renewcommand{\thefigure}{S\arabic{figure}}
\renewcommand{\theequation}{S\arabic{equation}}

\section{Theoretical Details}
\subsection{Electronic excitation efficiency}
The theoretical details are given in Ref.~\cite{Lu2012}. Here, for the sake of completeness, we re-produce the key results used in this work. 
In the extended wideband approximation, the nonequilibrium excitation of the vibrational mode is characterized by the excess noise spectrum\cite{Lu2012}
\begin{align}
    \nonumber\Delta \hat{\Pi}(\omega) = \frac{1}{2}\sum_{\sigma=\pm}(\chi^+-i\sigma \chi^-)(\hbar\omega+\sigma eV)\\
    \left[\coth\left(\frac{\hbar\omega+\sigma eV}{2k_BT}\right)-\coth\left(\frac{\hbar\omega}{2k_BT}\right)\right].
\end{align}
Here, $\chi^+$ and $\chi^-$ are matrices, whose size is the number of vibrational degrees of freedom. 
For single-molecular junctions (SMJs) considered here, $\chi^-$ is negligible, and we only need to consider the diagonal parts of $\chi^+$. For mode $n$, we have
\begin{align}
    \chi^+_n = \pi^{-1} {\rm Re}{\rm Tr}[M^nA_L(\mu_0)M^nA_R(\mu_0)]
\end{align}
characterizes the nonequilibrium excitation of the vibrational mode by electrons. The trace here is over electronic degrees of freedom.
Damping of the same mode $n$ is characterized by the corresponding diagonal matrix element of $\Gamma_e$
\[ \eta_n = \Gamma_{e,nn} .\]
Thus, the excitation efficiency is defined as
\begin{align}
    \gamma_n = \hbar\omega_n \chi^+_n/\eta_n.
\end{align}

\subsection{Power spectrum}
The kinetic energy power spectrum defined as
\begin{equation}
    C_{vv}(\omega) = \sum_i C_{v_i v_i}(\omega) = \sum_i \int dt C_{v_i v_i}(t) e^{i\omega t},
    \label{eq:cvv}
\end{equation}
where
\begin{equation}
    C_{v_i v_i}(t) = \langle v_i(t) v_i(0) \rangle = \frac{1}{T_0}\int dt' v_i(t'+t)v_i(t')
\end{equation}
is the velocity correlation function at steady state. The integration is over time range of length $T_0$ after the system reaches steady state. 
A cumulative integration of the power spectrum quantifies contribution of each frequency range to the total kinetic energy
\begin{equation}
    E_K (\omega) = \int_0^\omega d\omega' C_{vv}(\omega').
    \label{eq:kinetic}
\end{equation}

\section{Supporting Figures}
\begin{figure}[b]
	\centering
    \includegraphics[width=.8\columnwidth]{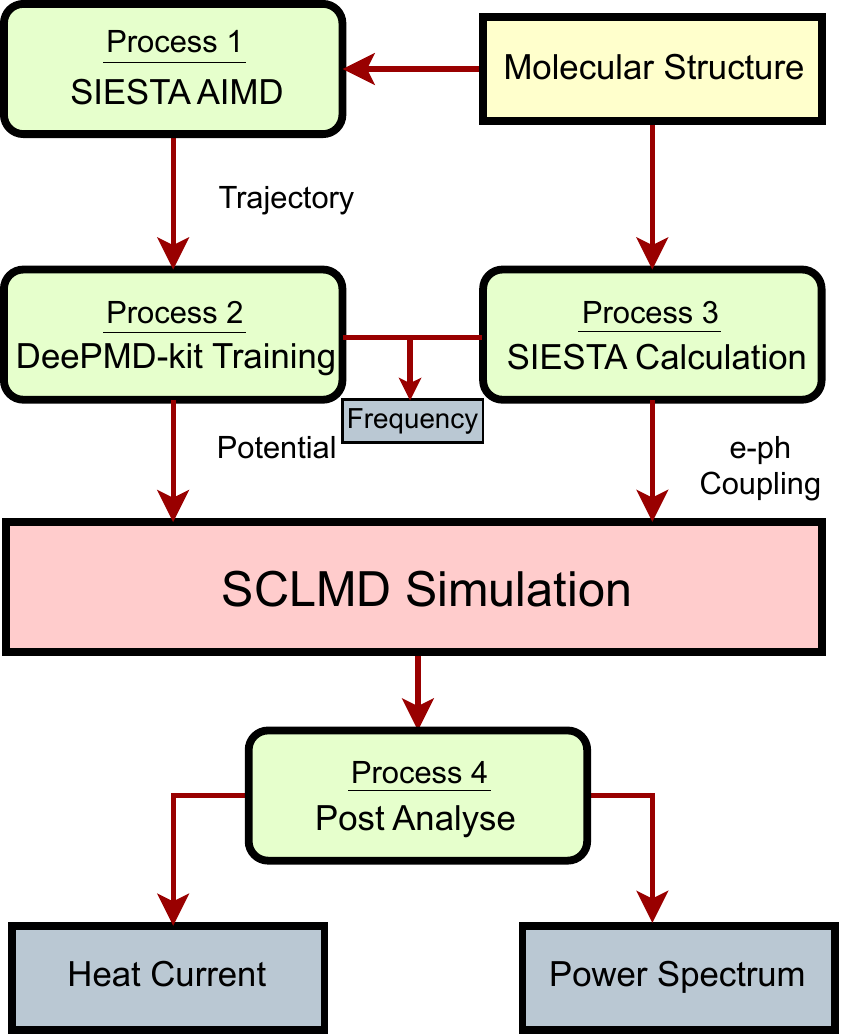}
    \caption{Work flow diagram. The red arrows present the workflow. After DeePMD-kit training of SIESTA molecular dynamics (MD) data and DFT (SIESTA-TranSIESTA-Inelastica toolkit) calculation of the electronic structure, vibrational spectrum, electron-vibration coupling, the parameters are passed to homemade MD package based on semi-classical Langevin equation to perform current-induced MD. The power spectrum and heat flux are obtained from resulting MD trajectories.}
    \label{fig:flow}
\end{figure}

\begin{figure}
    \centering
    \includegraphics[width=.8\columnwidth]{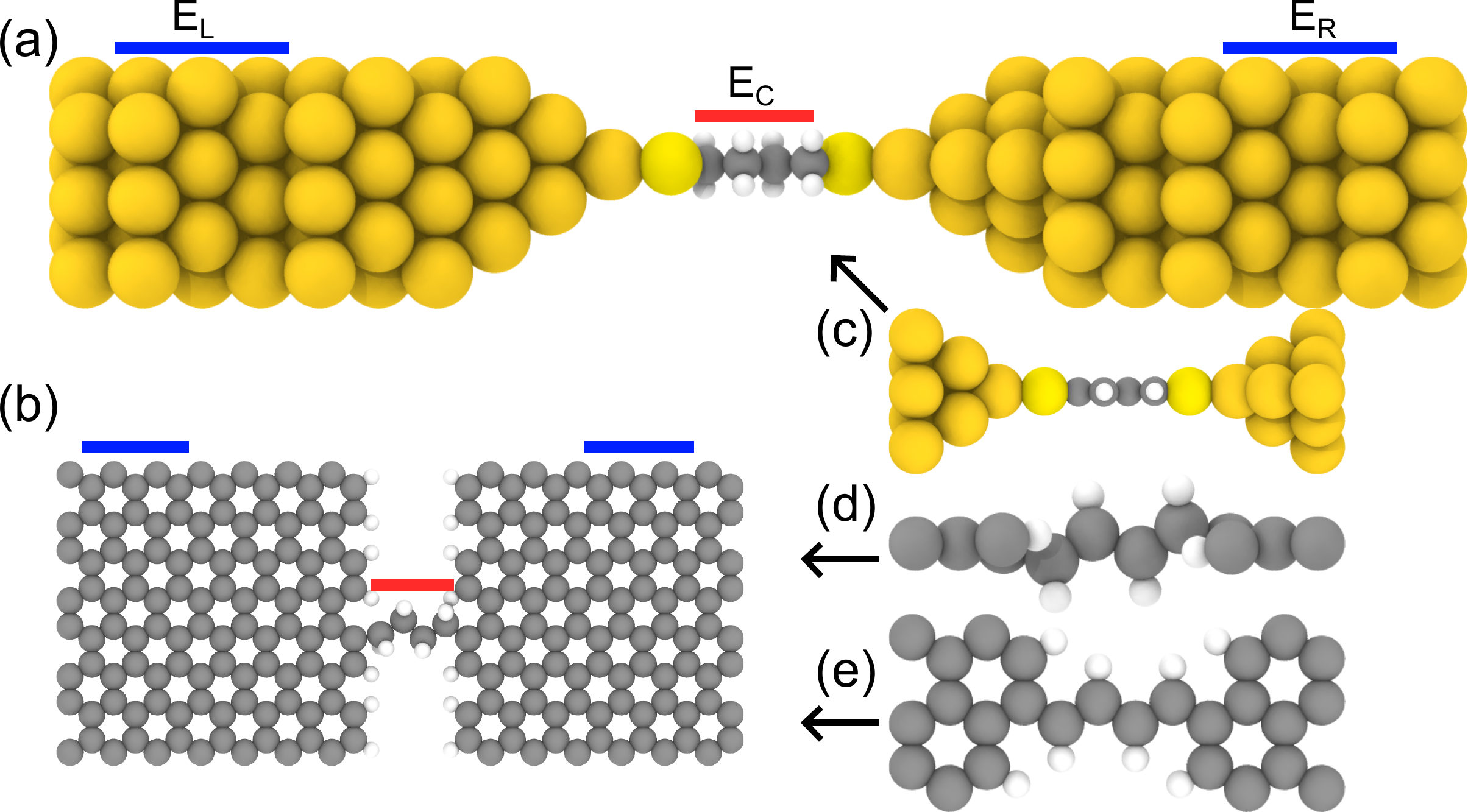}
    \caption{Structures of molecular junctions considered in this work. We consider two types of molecular structure with CH$_2$ (a, b) and CH (c, d, e) unit, respectively. The electrodes are either gold (a, c) or graphene (b, d, e). }
    \label{fig:str}
\end{figure}

\begin{figure}
	\centering
    \includegraphics[width=.8\columnwidth]{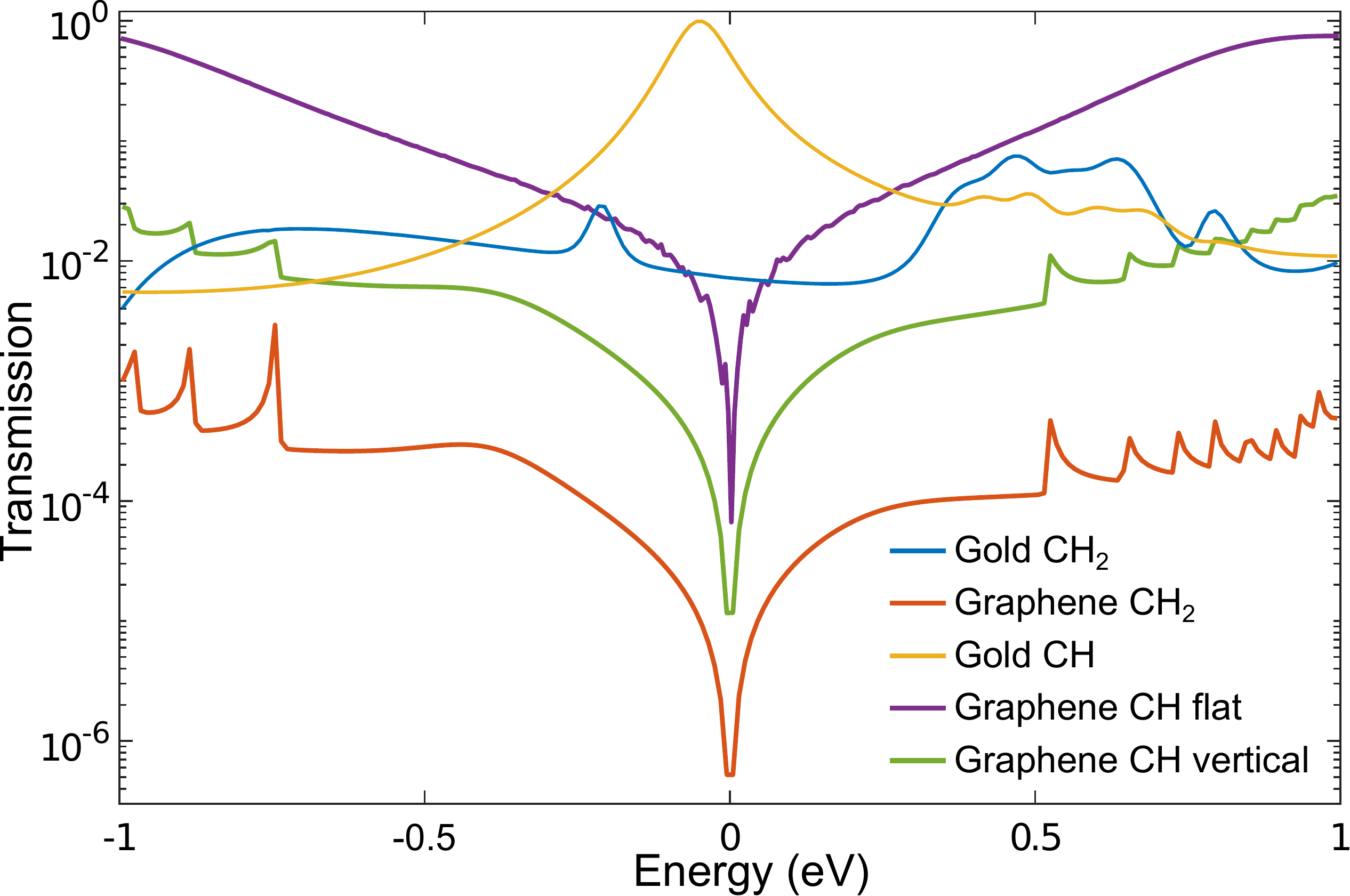}
    \caption{Electron transmission spectra of molecular junctions shown in Fig.~\ref{fig:str}.}
    \label{fig:trans}
\end{figure}

\begin{figure}
	\centering
    \includegraphics[width=.8\columnwidth]{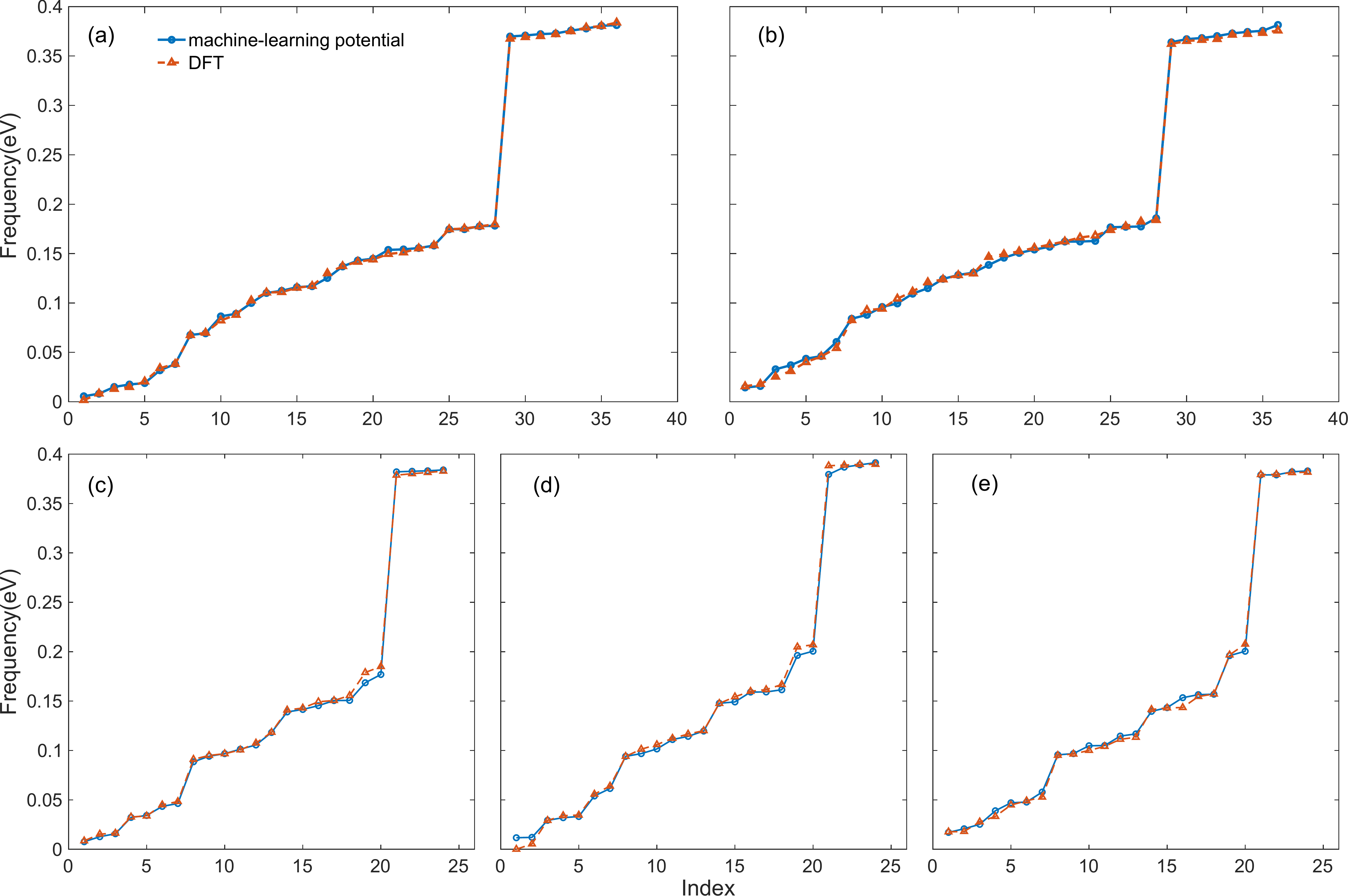}
    \label{fig:vib}
    \caption{Comparison of molecular vibration frequencies of molecular junctions shown in Fig. S2 from machine-learning potential and DFT. Blue circles are results from machine-learning potential and red triangles are from DFT.}
\end{figure}
\clearpage
\bibliography{ref1,ref,my,Heating,CurrentInducedForces}
\end{document}